\documentclass[prl,reprint,letterpaper,superscriptaddress,showpacs,amsmath,twocolumn,floatfix,longbibliography]{revtex4-1}
\RequirePackage{graphicx}
\usepackage{hyperref}
\newcommand{\D}{\mathrm d}

\newcommand{\I}{\mathrm i}
\renewcommand{\vec }{\mathbf }

\begin{document}

\title{On Dissipation Rate of Ocean Waves due to White Capping.}

\date{\today}

\author{A.\,O.~Korotkevich}
\email{alexkor@math.unm.edu}
\affiliation{Department of Mathematics and Statistics, MSC01 1115, 1 University of New Mexico, Albuquerque, NM 87131-0001, USA}
\affiliation{L.\,D.~Landau Institute for Theoretical Physics RAS, 2 Kosygin Str., Moscow, 119334, Russian Federation}

\author{A.\,O.~Prokofiev}
\email{alexpro@itp.ac.ru}
\affiliation{L.\,D.~Landau Institute for Theoretical Physics RAS, 2 Kosygin Str., Moscow, 119334, Russian Federation}

\author{V.\,E.~Zakharov}
\email{zakharov@math.arizona.edu}
\affiliation{Department of Mathematics, The University of Arizona, 617 N. Santa Rita Ave., P.O. Box 210089, Tucson, AZ 85721-0089, USA}
\affiliation{P.\,N.~Lebedev Physical Institute RAS, 53 Leninsky Prosp., GSP-1 Moscow, 119991, Russian Federation}
\affiliation{Novosibirsk State University, Novosibirsk, 630090, Russian Federation}


\begin{abstract}
We calculate the rate of ocean waves energy dissipation due to whitecapping by numerical simulation of deterministic phase resolving model for dynamics of ocean surface. Two independent numerical experiments are performed. First, we solve the $3D$ Hamiltonian equation that includes three- and four-wave interactions. This model is valid for moderate values of surface steepness only, $\mu < 0.09$. Then we solve the exact Euler equation for non-stationary potential flow of an ideal fluid with a free surface in $2D$ geometry. We use the conformal mapping of domain filled with fluid onto the lower half-plane. This model is applicable for arbitrary high levels of steepness. The results of both experiments are close. The whitecapping is the threshold process that takes place if the average steepness $\mu > \mu_{cr} \simeq 0.055$. The rate of energy dissipation grows dramatically with increasing of steepness. Comparison of our results with dissipation functions used in the operational models of wave forecasting shows that these models overestimate the rate of wave dissipation by order of magnitude for typical values of steepness.
\end{abstract}

\maketitle
\section{Introduction}
Formation of the white caps in the sea-waves crests is a well known physical phenomenon.
Its study is important for at least two reasons. White capping is a powerful mechanism
of waves energy dissipation, responsible for transfer of energy and momentum from wind
to water. Determination of the ``dissipation function'' --- the rate of energy and momentum
transport from atmosphere to water due to white capping in the wind-driven sea is the
necessary condition for designing of an efficient operational model for wave prediction.
During the last two decades the progress in improving of accuracy of acting operational
models (WAM3, WAM4, WAVEWATCH) was slow. On our opinion the main reason of this staggering
is the use of not well justified, if not just simply wrong, forms of the dissipation function.
Thus, proper parametrization of the dissipation function is the question of great practical
importance. At the moment at least three essentially different dissipation functions are used~\cite{Tolman2009}.
All of them are purely heuristic, none of them is justified by direct experimental or theoretical
foundations.

Study of white capping is also a question of serious theoretical significance. White capping
is a conspicuous example of the localized in space and time strongly nonlinear processes of
wave interaction. Another key class of such phenomena is the wave collapses like self-focusing
in nonlinear optics~\cite{SS1999} or Langmuir collapse in plasma~\cite{Zakharov1972}.
Wave collapses, like white capping, are also a mechanism for formation of hot spots of energy
dissipation, but further discussion of this subject is beyond the framework of the presented
article. We just mention that the theory of wave collapses is much better developed than the
theory of white capping~\cite{SS1999,Zakharov1972}.

From the mathematical point of view a collapse is formation of singularity of the basic equations
in a finite time. As a rule they are described by a self-similar solution of the basic equation~\cite{SS1999}.
This solution is at least a local attractor, thus collapses have a standard universal form.
The white capping is a more complicated and less studied phenomenon. A typical scenario of white
capping is formation of a narrow spray (``tongue'') on the crest of the breaking wave. This spray
rapidly becomes unstable, turbulent and decomposes to a cloud of drops. It is not clear how
universal is a form of a breaking wave. A theoretical study of white capping is enormously
difficult problem (e.g. see review~\cite{PZ2016}, recent paper~\cite{DN2016} proposed detailed
2D simulation of the initial stage of foam formation).

Nevertheless, well justified determination of the dissipation function is not a hopeless problem.
Any possible scenario of white capping starts from dramatical growth of the surface curvature in
some small localized area (for one of the mechanisms of foam formation see recent work~\cite{DN2016}).
One can move in the reference frame where this area is resting and
make a simple and plausible conjecture: all energy and momentum captured in the white capping
region will be rapidly dissipated. After accepting this conjecture we are not interested any more
in a detailed picture of the wave-breaking event. We just shall start to study this problem in
terms of Fourier transform of the surface shape. In other words, the following analogy can be useful:
if we throw a stone from the cliff, we know for sure that it will fall and details of stone rotation
during the way to the ground are not interesting for us.

Formation of zones of high curvature means generation of ``fat tails'' in the spatial spectra
of surface elevation. The conjecture formulated above means that all energy and momentum
supplied to these tails vigorously dissipate. This phenomenon cannot be studied in a framework
of the Hasselmann kinetic equation. To catch it we need to exploit any kind of dynamic
(i.e. phase resolving) equations, either exact or approximate. The dissipation can be introduced
in the model by addition of the artificial hyperviscosity to these equations. 
Such hyperviscose terms are acting only in the area of high enough
wave vectors and absorbing all energy supplied to this area. In virtue of energy conservation
in process of wave-wave interaction, the energy absorption in the white capping region is
exactly loss for the energy containing part of spectrum. This quantity can be measured in a
numerical simulation.

We realized this strategy in two independent and completely different massive numerical
experiments described below. We obtained very similar, almost coinciding results leading to
a fundamental conclusion --- white capping is a threshold phenomenon.
This is not a new idea. It was formulated by M.\,Banner and his collaborators~\cite{BT1998,BBY2000,SB2002}
almost two decades ago. In this paper we present a new argument in support of this idea,
obtained by straightforward numerical experiment.

In a wind driven sea the crucial role is played by a dimensionless parameter -- steepness. There are different
definitions of steepness. If $\eta(\vec r, t)$ is the shape of the surface (deviation from the unperturbed
state), its dispersion $\sigma$ is defined as $\sigma^2=\langle \eta^2 \rangle$. This quantity in oceanography
sometimes called ``energy''. The natural ``physical'' definition of steepness $\mu$ is
\begin{equation}
\mu^2 = \langle |\vec \nabla \eta|^2 \rangle,
\end{equation}
here  $\vec \nabla$ is a gradient computed in the plane of the fluid surface, usually $XY$-plane. This definition
corresponds to an average slope of the surface. However,
the ``physical'' steepness is hard to measure directly in a field experiment. For this reason it is
convenient to introduce the ``oceanographic'' steepness
\begin{equation}
S^2 = k_p^2 \sigma^2,
\end{equation}
here $k_p$ is a wavenumber of the spectral peak. For pure monochromatic waves these two definitions coincide.
For real energy spectra, which in ocean have powerlike ``tails'', they will differ: ``physical'' steepness will be
higher. In our experiments spectra are relatively close to monochromatic ones, so we will not distinguish
between these two definitions and use notation $\mu$ everywhere.

In accordance with already formulated concepts of M.\,Banner and his collaborators~\cite{BT1998,BBY2000,SB2002},
we have found that white capping (wave breaking) is the critical phenomenon with critical
value of steepness $\mu_{cr}\simeq 0.056$. If $\mu < \mu_{cr}$ white capping is practically absent.
If $\mu \ge \mu_{cr}$, dissipation function $S_{diss}$ is the fast growing function on
$\mu - \mu_{cr}$.

This dependence of the dissipation function on average steepness $\mu_p$ differs dramatically
from widely accepted parametrization $S_{diss}\simeq \mu_p^4$~\cite{KCDHHJ1994}.
We would like to stress once more, that our results perfectly corroborate with experimental observations made in ocean and on
Lake Victoria by Banner, Babanin and Young~\cite{BT1998,BBY2000,SB2002}. They did not measure energy dissipation
due to white capping but they studied probability of wave breaking event as a function of
the average steepness. They obtained the same result --- white capping is the critical phenomenon
and the value of $\mu_{cr}$ was practically the same. A probability of wave breaking is almost zero
if $\mu_p < \mu_{cr}$ and grows dramatically with increasing of $\mu_p - \mu_{cr}$. This is exactly
what we observed in our numerical experiments.

Our numerical results lead to a very important practical conclusion: existing operational wave
forecasting models essentially overestimate the role of the white capping dissipation, especially
for ``mature'' or relatively ``old'' sea, which is characterized by low $\mu_p$. The dissipation
function must be seriously diminished. We realize that it means total reexamination of the balance
equation in the equilibrium area of spectra and choosing of a more realistic model of the
wind input term $S_{in}$. However, these questions are beyond the scope of this article.

Preliminary results of this paper were published in~\cite{ZKP2009}. Recently another numerical experiment
using significantly simplified version of dynamical equations was published in~\cite{DKZ2015}. Its results
although based on different model, completely confirmed the results of the present paper as well as our main conclusion:
dissipation functions used in operational models overestimate contribution of white capping to waves energy
absorption and must be reexamined.

\section{Basic models.}
We study the potential flow of an ideal inviscid incompressible fluid with the velocity potential
$\phi=\phi(x,y,z;t)$, satisfying the Laplace equation
\begin{equation}
\label{Laplace}
\Delta \phi = 0.
\end{equation}
The fluid is deep, thus we solve equation~(\ref{Laplace}) inside the domain
$-\infty < z < \eta(\vec r, t)$, $\vec r = (x,y)$ --- coordinate on the still fluid.
Let $\psi=\phi|_{z=\eta}$ -- velocity potential evaluated on the surface. Fixation of
$\eta=\eta(\vec r, t)$, $\psi=\psi(\vec r, t)$ together with condition $\phi_z\rightarrow 0$,
$z\rightarrow -\infty$ define the Dirichlet-Neumann uniquely resolvable boundary condition
for the Laplace equation. As it was shown in~\cite{Zakharov1968}, $\eta$ and $\psi$ obey the
Hamiltonian equations:
\begin{equation}
\label{Hamiltonian_equations}
\frac{\partial \eta}{\partial t} = \frac{\delta H}{\delta \psi}, \;\;\;\;
\frac{\partial \psi}{\partial t} = - \frac{\delta H}{\delta \eta},
\end{equation}
where $H$ is the Hamiltonian of the system
$$
H = \frac{1}{2} \int_{-\infty}^{+\infty} \D x \D y \left(
g\eta^2 + 
\int_{-\infty}^{\eta} |\nabla \phi|^2 \D z
\right),
$$
Unfortunately $H$ cannot be written in the close form as a functional of $\eta$ and $\psi$.
However one can limit Hamiltonian by first three terms of expansion in powers of $\eta$ and $\psi$~\cite{Zakharov1968}
\begin{equation}
\label{Hamiltonian}
\begin{array}{l}
\displaystyle
H = H_0 + H_1 + H_2 + ...,\\
\displaystyle
H_0 = \frac{1}{2}\int\left( g \eta^2 + \psi \hat k  \psi \right) \D x \D y,\\
\displaystyle
H_1 =  \frac{1}{2}\int\eta\left[ |\nabla \psi|^2 - (\hat k \psi)^2 \right] \D x \D y,\\
\displaystyle
H_2 = \frac{1}{2}\int\eta (\hat k \psi) \left[ \hat k (\eta (\hat k \psi)) + \eta\Delta\psi \right] \D x \D y.
\end{array}
\end{equation}
Here $\hat k$ is a linear integral operator 
$\left(\hat k =\sqrt{-\Delta}\right)$, such that in $k$-space it corresponds to
multiplication of Fourier harmonics ($\psi_{\vec k} = \frac{1}{2\pi} \int \psi_{\vec r} e^{i {\vec k} {\vec r}} \D x \D y$)
by $k=\sqrt{k_{x}^2 + k_{y}^2}$. For gravity waves this
reduced Hamiltonian describes four-wave interaction.

Then dynamical equations (\ref{Hamiltonian_equations}) acquire the form
\begin{eqnarray}
\dot \eta = \hat k  \psi - (\nabla (\eta \nabla \psi)) - \hat k  [\eta \hat k  \psi] +\nonumber\\
+ \hat k (\eta \hat k  [\eta \hat k  \psi]) + \frac{1}{2} \Delta [\eta^2 \hat k \psi] + 
\frac{1}{2} \hat k [\eta^2 \Delta\psi] - \widehat F^{-1} [\gamma_k \eta_k],\nonumber\\
\dot \psi = - g\eta - \frac{1}{2}\left[ (\nabla \psi)^2 - (\hat k \psi)^2 \right] - \label{eta_psi_system}\\
- [\hat k  \psi] \hat k  [\eta \hat k  \psi] - [\eta \hat k  \psi]\Delta\psi - \widehat F^{-1} [\gamma_k \psi_k].\nonumber
\end{eqnarray}
Here dot means time-derivative, $\Delta=\nabla^2$ --- Laplace operator,
$\widehat F^{-1}$ is an inverse Fourier transform, $\gamma_k$ is a
dissipation rate (according to work~\cite{DDZ2008} it has to be included in both equations), which
corresponds to viscosity at small scales. The detailed description of used numerical algorithm
for solving system~\eqref{eta_psi_system} was published in our recent paper~\cite{KDZ2016}.

Expansion~(\ref{Hamiltonian}) is valid for full 3D problem. In the simpler 2D case dependence on $y$
drops out and one can introduce complex coordinate
$Z = x + \I z$,
on the physical plane and perform conformal mapping onto the lower half plane
$W = u + \I v$.
The conformal mapping is completely defined by the Jacobian
$R = W_Z = \frac{1}{Z_W}$.
The fluid flow is defined by the complex velocity potential
$\Phi = \Psi + \I\widehat H \Psi$.
Here $\widehat H$ --- the Hilbert transform:
$\widehat H \Psi = \frac{1}{\pi}v.p.\int_{-\infty}^{+\infty}\frac{\Psi(s)}{s-u}\D s.$
The complex velocity is given by equation
$V = \I\frac{\partial \Phi}{\partial z}$.
Both $R$ and $\Phi$ are analytic functions in the lower half-plane $v < 0$.
Dyachenko has shown that $R$ and $V$ obey the following
``Dyachenko equations''~\cite{Dyachenko2001}:
\begin{equation}
\frac{\partial R}{\partial t} = \I (UR' - RU'),\;
\frac{\partial V}{\partial t} = \I (UV' - RB') + g(R-1).\label{DyachenkoEqs}
\end{equation}
Here $U = P^{-}(R\bar V + \bar R V)$ and $B = P^{-}(V\bar V)$.
$P^{-} = 1/2 (1 + \I \widehat H)$ --- operator of projection to the lower half-plane.
We must stress that~(\ref{DyachenkoEqs}) are {\bf exact} equations, completely equivalent
to the primordial Euler equations. In the present paper we used numerical algorithm analogous to the one used in~\cite{ZDP2006}.

\section{Experimental results.}
\subsection{3D experiment.}
We simulate primordial dynamical equations~(\ref{eta_psi_system})
in a $2\pi\times 2\pi$ periodic box with spectral resolution $512\times4096$.
We did not impose any external forcing but studied decay of initially created wave turbulence, which is almost
unidirectional swell and can be simulated in such
rectangular spectral domain due to strong anisotropy of the wave field.
We used the following damping parameters:
\begin{equation}
\label{Pseudo_Viscous_Damping}
\gamma_k = \left\{
\begin{array}{l}
\displaystyle
0, k < k_d,\\
\displaystyle
- \gamma (k - k_d)^2, k \ge k_d,\\
\end{array}
\right.\;
\begin{array}{l}
\displaystyle
k_d = 1024,\\
\displaystyle
\gamma = 2.86\times 10^{-3}.
\end{array}
\end{equation}
Gravity acceleration $g = 1$, time step $\Delta t = 4.23\times 10^{-4}$.

As initial conditions we used Gauss-shaped spectrum, centered at $\vec k_0 = (0; 100)$ with width $D_i = 30$.
All harmonics amplitudes were random Gaussian values with the following average
\begin{equation}
\label{Dynamic_initial_conditions}
\begin{array}{l}
\displaystyle
\left\{
\begin{array}{l}
\displaystyle
|a_{\vec k}| = A_i \exp \left(- \frac{1}{2}\frac{\left|\vec k - \vec
k_0\right|^2}{D_i^2}\right),
\left|\vec k - \vec k_0\right| \le 2D_i,\\
\displaystyle
|a_{\vec k}| = 10^{-12}, \left|\vec k - \vec k_0\right| > 2D_i.
\end{array}
\right.
\end{array}
\end{equation}
Phases were random numbers uniformly distributed in the interval $[0;2\pi)$.
The value of $A_i$ was varying in order to provide desired initial average steepness
$\mu=\sqrt{\langle|\nabla\eta(\vec r)|^2\rangle}$, which was in the interval from 0.05 to 0.09.

During the experiment we observed dynamics of initial spectral distribution during
time $100T_0$, here $T_0$ is the period of the wave corresponding to spectral maximum.
This time is approximately 10 nonlinear times and we suppose that at this time initial
spectral distribution is relaxed close enough to quasistationary 
self-similar state. This means formation of powerlike Kolmogorov-Zakharov (KZ) spectral tails~\cite{DKZ2003grav,DKZ2004},
which are routinely observed in field and laboratory experiments, e.g. see~\cite{Donelan1985}.

We have approximately one decade inertial interval in wave numbers. It lets us to observe formation of a
long spectral tail and to secure
our spectral maximum from the direct influence of the dissipation region. We would like to stress,
that this is not far from reality. In the wind-driven sea the inertial interval for gravity waves is
posed approximately in the following range (see~\cite{Donelan1985})
$$
\omega_p < \omega < 3.5\omega_p,
$$
here $\omega_p$ is a spectrum peak circular frequency corresponding to the wave number $k_p$.
This spectral band is broad enough to provide a possibility
for formation of sharp crests, where local steepness is much higher then average which could lead to
breakdown of our dynamical equations~(\ref{eta_psi_system}), derived in assumption that we can expand Hamiltonian in powers of steepness,
so it gives a limitation on the value of the initial average steepness. We also have to keep average steepness
high enough to secure four wave interaction from been stopped by discreteness of the wavenumbers
grid~\cite{ZKPD2005}.
To achieve a balance between these two requirements was the main problem during the experiment.

In the process of the experiment we calculated the inverse normed time of wave action derivative:
$$
\gamma_{diss} = \frac{N'}{\omega_0 N} = 2\frac{N(t + 10T_0) - N(t)}{10T_0 \omega_0 (N(t + 10T_0) + N(t))},
$$
here $\omega_0 = \sqrt{g k_0}$, which was calculated at three different moments of time $t=70T_0$,
$t=80T_0$, and $t=90T_0$. Because initial wave field was stochastic in the first approximation
we can consider these three results as three experiments with random initial phase and amplitude, or
three independent cases in stochastic ensemble. The results of this experiment are presented in Figure~\ref{Gamma100_3D_2D}.

Let us mention that at $\omega\simeq 3.5\omega_p$ the KZ spectrum tails $\sim \omega^{-4}$ transit at least at some conditions
to the Phillips spectrum $\sim \omega^{-5}$~\cite{Korotkevich2008PRL,Korotkevich2012MCS}. This transition needs to be studied more carefully. Some preliminary results can be found
in~\cite{NZ1992,ZB2012}.

\subsection{2D experiment}
In this experiment we solved Dyachenko equations~\eqref{DyachenkoEqs} in the periodic spatial domain of length $2\pi$ with spectral resolution of 8192 harmonics. We replaced in equations for $R$ and $V$ time derivative by:
\begin{equation}
\frac{\partial}{\partial t} := \frac{\partial}{\partial t} - \gamma_p(k) + 2.0\times10^{-12}k^{4}.
\end{equation}
Here $\gamma_p(k)$ is symmetric in $k$-space pumping function concentrated near $k=k_0=35$.
As initial data we chose $R=1$, $V$ - is the white noise consisting of harmonica with random phase and amplitude $10^{-20}$.

\begin{figure}[htb]
\includegraphics[width=3.5in]{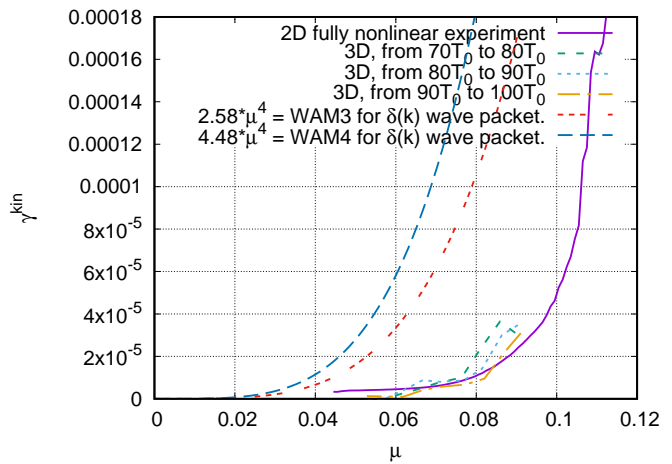}
\caption{\label{Gamma100_3D_2D} Dissipation term $\gamma_{diss}$ as a function of average surface steepness. Both 3D and 2D experiments are presented.}
\end{figure}

On the initial stage of experiment pumping forms quasimonochromatic standing wave of a very high steepness
$\mu=\sqrt{\langle(\eta'(x))^2\rangle}\simeq 0.2$.
Then we switched off the pumping and measured
$\gamma_{diss}=\frac{N'}{\omega_p N}$,
where $\omega_p$ is a frequency at $k_p$.
 The idea was to reach stepnesses which are way beyond the range of applicability of truncated 3D
model and observe the dissipation there, eventually bringing the solution to the values of steepness manageable by 3D experiment.
We observed fast dissipation which slowed down dramatically when steepness achieved value $\mu\simeq 0.1$.
The dissipation rate for small steepnesses is presented in Figure~\ref{Gamma100_3D_2D}
together with the results of the 3D experiment.

\begin{figure}[htb]
\includegraphics[width=3.5in]{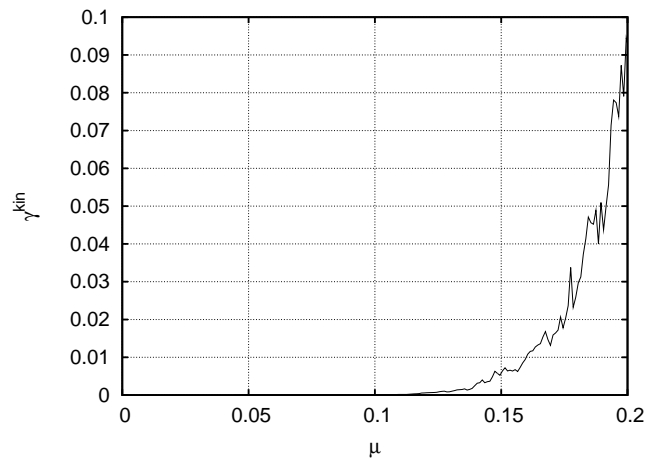}
\caption{\label{Gamma100_2D} Dissipation term $\gamma_{diss}$ as a function of average surface steepness $\mu=\sqrt{\langle|\eta'(x)|^2\rangle}$.}
\end{figure}

Dissipation rate for higher values of steepness is given in Figure~\ref{Gamma100_2D}
One can see that steep waves loose energy extremely fast, during few wave periods.

\subsection{Comparison with wave forecasting models}
Widely used  wave forecasting models WAM3 and WAM4 utilize the following dissipation functions depending on wave number $k$:
\begin{eqnarray}
\label{WAMdissipation}
\gamma_{\vec{k}} = C_{ds}  \tilde{\omega} \frac{k}{\tilde{k}} \left((1-\delta)+
\delta\frac{k}{\tilde{k}}\right)\left(\frac{S}{S_{PM}}\right)^p
\end{eqnarray}
where $k$ and $\omega$ are the wave number and frequency, tilde denotes mean value;
$C_{ds}$, $\delta$ and $p$
are tunable coefficients; $S=k_p\sigma$ is the ``oceanographic'' steepness;
$S_{PM}=(3.02\times 10^{-3})^{1/2}$
is the value of $\tilde{S}$ for the Pierson-Moscowitz spectrum.
The values of the tunable coefficients for the {\it WAM3} case are:
\begin{equation}
C_{ds} = 2.35\times 10^{-5},\,\,\,\delta=0,\,\,\,p=4
\end{equation}
and for the {\it WAM4} case are:
\begin{equation}
C_{ds} = 4.09\times 10^{-5},\,\,\,\delta=0.5,\,\,\,p=4
\end{equation}
As long as we are interested only in rough estimate we can identify $S\simeq\mu$ and consider the spectrum as completely concentrated near the spectral
maximum $k=\tilde k$, $\omega=\tilde \omega$. As a result we end up with following normalized dissipation rates $\gamma_{diss}\simeq 2.58\mu^4$ for WAM3
and $\gamma_{diss}\simeq 4.48\mu^4$ for WAM4. Corresponding curves are plotted in Figure~\ref{Gamma100_3D_2D}. One can see that for typical wind driven
sea steepness $\mu\simeq 0.06$ they overestimate the dissipation due to white capping by almost order of magnitude.

We must stress that we observed the ``threshold type'' onset of dissipation only in the case when spectral peak scale $k_p$
and beginning of dissipation scale $k_d$ are separated far enough in $k$-space. If they are relatively close to each other (say, $k_d \sim 3k_p$ or $k_d \sim 5k_p$, we performed both these experiments) we observe a smooth powerlike dependence of dissipation from steepness, similar to what was
offered in WAM3 model. This is due to drain of energy from the spectral peak through slave harmonics in dissipation range. Such a dependence can be
obtained analytically ($p=4$ for the case $k_d \sim 3k_p$). It has to be noted that in the real wind-driven sea these scales are separated
by at least an order of magnitude. 

\section{Discussion and results}
One could ask a question: ``Why authors used these two models with such different initial conditions?''. The answer is simple. If the mechanism of
dissipation which we proposed (domination of generation of multiple harmonics, when local singularities start to appear, aided by direct cascade energy transfer) is universal, then results in this two models (just a reminder, four-wave resonant interactions are absent in 2D case, which means different
nonlinear interaction physics; while multiple harmonics generation is the same) will be not necessarily the same but close. Also, using fully nonlinear
2D model we can reach range of steepnesses corresponding to rough sea. As we expected, simulations suggest that the mechanism is universal.
Results are discussed in details below.

First of all, both our experiments clearly show that dissipation functions widely used in operational wave forecasting models dramatically overestimate
the role of white capping in the energy balance of wind-driven sea. Another, less direct reasons, in virtue of this statement were formulated
in paper~\cite{PZ2016}.

Comparison of 3D and 2D experiments for ``realistic'' values of steepness
$0.06 < \mu < 0.08$ 
shows that in 3D case the dissipation is somewhat stronger. This fact has a natural explanation.
As it was mentioned above, there are two completely different mechanisms of energy transfer
to the small scale region of spectrum where waves dissipate. One is the Kolmogorov-type flux of energy due to four-wave resonant processes
(see, for instance~\cite{Zakharov2010}). Second is the direct formation of white caps on crests of waves. Strong local nonlinearity leads to
fast growth of multiple harmonics, which is much faster than energy cascade. In 2D geometry only second mechanism is working.
We are not able to compare efficiency of these two mechanisms for high steepnesses ($\mu > 0.08$) because for such steep waves any type of weakly
nonlinear models fail. Meanwhile, 2D-experiments show that the white capping dissipation of steep waves (since $\mu\simeq 0.1$) is an extremely powerful
process, and the waves close to the critical Stokes wave ($\mu > 0.3$) just cannot exist in reality. They break up very rapidly. The mechanism
of regularization of such process through formation of limiting capillary waves on the crest was recently modeled and described in~\cite{DN2016}.

Our 3D experiments confirmed the idea of Banner and his co-authors~\cite{BBY2000} about ``threshold'' character of white capping. There are several
experimental evidences in support of comparison of the wave breaking phenomenon with second-order phase transition (e.g. see~\cite{NZ1992}).
The threshold level $\mu_{cr}\simeq 0.055$ coincides with experimental data~\cite{PZ2016,BBY2000} very well.
In 2D experiment results are less evident. As it is seen in the Figure~\ref{Gamma100_3D_2D}, some dissipation remains even for waves of small steepness.
This remainder of dissipation does not depend of steepness at all. This is a numerical artifact.
We were not able to stabilize our numerical scheme without introducing of artificial hyperviscosity in the whole
$k$-space. Recent experiments~\cite{DKZ2015} performed in a framework of a less accurate (and more robust) model of 2D waves, support the threshold
nature of white capping.

Our next task is formulation of a dissipation function, more realistic than~\eqref{WAMdissipation}. The first step in this direction was made in~\cite{ZB2012}. Here we propose as a first preliminary results nonlinear least square (Marquardt-Levenberg algorithm implementation in Gnuplot~\cite{Gnuplot})
fits of two functions. Probably the most interesting is a region of relatively small steepness, corresponding to mature or not so rough sea
($0.055 \le \mu \le 0.1$). We tried two types of functions, exponential:
\begin{equation}
\gamma_k^{lowExp}(\mu) = (1.248\times 10^{-7})\exp(59.40\mu),
\end{equation}
and polynomial in terms of difference $\mu - \mu_{cr}$:
\begin{equation}
\gamma_k^{lowPoly}(\mu) = (1.040\times 10^{-2})|\mu - 0.055|^{1.756}.
\end{equation}
For stepnesses corresponding to rough sea (up to $\mu = 0.2$) we obtained different fits. Exponential function yielded the following dependence:
\begin{equation}
\gamma_k^{highExp}(\mu) = (8.679\times 10^{-7})\exp(58.00\mu),
\end{equation}
and polynomial fit gave this result:
\begin{equation}
\gamma_k^{highPoly}(\mu) = (1.468\times 10^{5})|\mu - 0.055|^{7.398}.
\end{equation}
It should be stressed, that this choice of functions is kind of random. We add as a supplementary material to this paper the measured
dependencies of dissipation on steepness for every experiment, so anybody can try to choose any function and fit it using any ready package available.

Having said that, let us describe the format and the files in supplementary materials. All of them are plain text files which contain two columns:
the first one is average steepness and the second one is measured value of dissipation rate $\gamma$. File {\tt mu\_gamma\_3D.data} contains
data points from 3D experiments. For every experiment we averaged three values obtained from time intervals $80-70 T_0$, $90-80 T_0$, and
$100-90 T_0$ as it is described in corresponding section above. Both values of steepness and dissipation rate were averaged. File {\tt mu\_gamma\_2D.data}
contains data points from 2D experiments. One can see that for values of steepness $\mu < 0.065$ artificial viscosity affect the data as was discussed
in details above. This is why we decided to glue together both sets and use all data points from 3D case together with data points for higher steepness
from 2D case for purposes of fitting some functions to the dependence. This combined data set is given as a file {\tt mu\_gamma\_2D\_3D.data}. We hope
that these initial results can be used by community in order to improve dissipation functions currently used in different wave forecasting models.

\begin{acknowledgments}
In conclusion VZ would like to express gratitude for support to the Russian Scientific Foundation grant 14-05-00479. AK would like to acknowledge
support by National Science Foundation grant OCE 1131791. Analysis of the experimental data was performed by AK with support of Russian Scientific Foundation
grant 14-22-00259.
\end{acknowledgments}
%
\end{document}